\documentclass[12pt]{article}

\def\a{\alpha}\def\e{\epsilon}
\def\g{\gamma}
\def\l{\lambda}\def\m{\mu}\def\n{\nu}\def
\p{\pi}\def\r{\rho}
\def\y{\eta}

\def\L{\Lambda}

\def\na{\nabla}
\def\mo{{-1}}\def\ha{{1\over 2}}

\def\mn{{\mu\nu}}

\def\PL#1{Phys.\ Lett.\ {\bf#1}}
\def\PRL#1{Phys.\ Rev.\ Lett.\ {\bf#1}}
\def\PR#1{Phys.\ Rev.\ {\bf#1}}\def\CQG#1{Class.\ Quantum Grav.\ {\bf#1}}
\def\NP#1{Nucl.\ Phys.\ {\bf#1}}
\def\JMP#1{J.\ Math.\ Phys.\ {\bf#1}}

\def\grq#1{{\tt gr-qc #1}}\def\hep#1{{\tt hep-th #1}}

\def\Lh{{\L\over2}}
\def\er{{\cal R}}
\def\se{{\rm if}}
\def\mn{{\mu\nu}}
\def\na{\nabla}
\def\e{{\rm \, e}}
\def\rF{{\rm \, F}}
\def\sh{{\rm \, sinh\, }}
\def\ch{{\rm \, cosh\, }}
\def\th{{\rm \, tanh\, }}
\def\mo{{-1}}
\def\be{\begin{equation}}
\def\ee{\end{equation}}
\def\lb{\label}
\def\bea{\begin{eqnarray}}
\def\eea{\end{eqnarray}}
\def\nn{\nonumber}
\def\bd{\begin{displaymath}}
\def\ed{\end{displaymath}}

\begin{document}
\begin{titlepage}
\begin{flushright}
INFNCA-TH0202 \\
\end{flushright}
\vspace{.3cm}
\begin{center}
\renewcommand{\thefootnote}{\fnsymbol{footnote}}
{\Large \bf Cosmology of the Jackiw-Teitelboim model}
\vfill
{\large \bf {M.~Cadoni$^{1,a}$\footnote{email: cadoni@cagliari.infn.it}
and
S.~Mignemi$^{2,a}$\footnote{email: smignemi@unica.it}}}\\
\renewcommand{\thefootnote}{\arabic{footnote}}
\setcounter{footnote}{0}
\vfill
{\small
$^1$ Universit\`a di Cagliari, Dipartimento di Fisica,\\
Cittadella Universitaria, 09042 Monserrato, Italy\\
\vspace*{0.4cm}
$^2$ Universit\`a di Cagliari, Dipartimento di Matematica,\\
Viale Merello 92, 09123 Cagliari, Italy\\
\vspace*{0.4cm}
$^a$ INFN, Sezione di Cagliari\\
}
\end{center}
\vfill
\centerline{\bf Abstract}
\vfill

We investigate the cosmology of the two-dimensional
Jackiw-Teitelboim model.  Since the coupling of matter with gravitation
is not defined uniquely, we consider two possible choices.
The dilaton field plays an important role in the discussion of
the properties of the solutions. In particular, the possibility of
universes having a finite initial size emerges.

\vfill
\end{titlepage}

\section{Introduction}

Models of gravity in two dimensions have been largely studied in
recent years as toy models addressing issues that are too complex
to be faced directly in four dimensions.
However, the dynamics of two-dimensional gravity is rather
different from its four-dimensional counterpart, since the
Einstein-Hilbert action is a topological
invariant in two dimensions and hence gives rise to trivial field
equations. In order to derive the field equations from an
action principle it is then necessary either to resort
to higher-derivative theories \cite{Sc} or to
introduce an auxiliary scalar field $\y$ (which in the following
will be called dilaton) \cite{Ja}. This field
may be interpreted as the inverse of a spacetime-dependent gravitational
coupling constant, and hence, as has been remarked
in \cite{CM}, cannot be ignored in the discussion of the spacetime
structure. In particular, its zeroes should be regarded as true
physical singularities, since the gravitational force blows up there.

One of the main topics in the theory of gravitation is the
study of cosmological models. Recently this subject has received a
special attention also in the context of the investigation of gravitational
entropy bounds \cite {entbound}, and their relation to the holographic
principle \cite{tH}.
In \cite{CM2} this problem has been addressed in a two-dimensional
setting, using a Jackiw-Teitelboim action.
It appears therefore useful to examine more closely the cosmology
of two-dimensional models.

Actually, classical two-dimensional cosmology, which also emerges as a
limiting case of string cosmology \cite{mueller}, has been studied in
detail only in the context of a model where the curvature scalar
is  proportional to the trace of the energy-momentum
tensor\footnote{Curiously, more attention has been devoted to
two-dimensional quantum cosmology \cite{Li}.} \cite{Ma}.
Although this model can be derived from a dilaton-gravity action \cite{MM},
the role of the dilaton field has been neglected in these investigations.
However, in our opinion its role is essential in the interpretation of
the theory, since, as remarked above, in a cosmological context the
dilaton can be considered as a time-varying Newton constant.

In the following we shall consider the simplest dilaton-gravity
model in two dimensions, namely that of Jackiw and Teitelboim \cite{Ja,Te}.
Its specificity is that its action does not contain any
kinetic energy term for the dilaton. Nevertheless, it can be related to
a large class of equivalent models by conformal transformations
\cite{MS}.

The JT action for gravity coupled to matter reads:
\be\lb{act}
I=\int d^2x\sqrt{-g}\left(\eta{\er-\L\over16\p G}+L_M\right),
\ee
where $\er$ is the curvature scalar, $\L$ is a cosmological
constant and $G$ is the gravitational coupling constant, which is
dimensionless and may be absorbed in a redefinition of $\y$.
$L_M$ is the action of two-dimensional matter. In analogy
with the four-dimensional case, this can be
taken to be proportional to $-\r$, with  $\r$ the mass density.
However, the coupling of the matter with the dilaton $\y$ is not
fixed a priori. In fact, one may choose $L_M=-\y^\a\r$, for any
$\a$, giving rise to a large variety of inequivalent models.
In the following we shall consider the two simplest possibilities:
$\a=0$ (minimal coupling) and $\a=1$ (conformal coupling). These are
the most interesting for physics since are the closest to those
employed in higher dimensions.

\section{Minimally coupled matter}
In this section, we consider the case of minimally coupled matter.
This is the most straightforward generalization of the higher
dimensional formalism. However, as we shall see, the metric decouples
from matter in this model.

The field equations can be obtained by varying the action
(\ref{act}), with $L_M=-\r$. As in general relativity, in
order to obtain the correct energy-momentum tensor
the matter action must be subjected to a constrained variation
\cite{DC}.
The field equation then read
\bea
&&\er=\L,\lb{e1}\\
&&-(\na_\mu\na_\nu-g_\mn\na^2)\eta+\Lh g_\mn\eta=8\pi GT_\mn,\lb{e2}
\eea
with
$T_\mn=pg_\mn+(\rho+p)u_\mu u_\nu$. The conservation law for the
energy-momentum tensor, $\na^\m T_{\m\n}=0$, can
be obtained combining the covariant derivative of (\ref{e2}) with
(\ref{e1}) and recalling that in two dimensions
$[\na^\m,\na_\n]A_\m=\ha\er A_\n$, for any $A_\m$.

It is evident from the field equations that in this model the only
dependence on the matter content is through the dilaton, while the
metric function is unaffected
and depends only on the value of the cosmological constant.
This is in accordance with our interpretation of the dilaton as a
fundamental field of the theory.

We look for a solution of the form
\be\lb{ans}
ds^2=-dt^2+R^2(t)dr^2,\qquad\eta=\eta(t),
\ee
The coordinate $r$ can be either compact, $0\le r\le2\p$, or
non-compact. This choice does not affect the field equations,
that take the form
\bea
&&{\ddot R\over R}=\Lh,\lb{feq1}\\
&&{\dot R\over R}\dot\eta=\Lh\eta+8\pi G\rho,\lb{feq2}\\
&&\ddot\eta=\Lh\eta-8\p Gp.\lb{feq3}
\eea
Eq. (\ref{feq1}) admits the first integral
\be\lb{feq1a}
\dot R^2-\Lh R^2=a,
\ee
that can be immediately integrated to yield $R$.
Only two of the field equations are independent. In fact,
differentiating (\ref{feq2}) and combining with (\ref{feq1}) and
(\ref{feq3}), one obtains the energy-momentum conservation law
\be\lb{cons}
\dot\rho=-(p+\rho)\dot R/R.
\ee

For a perfect fluid, the equation of state is $p=\gamma\rho$,
with $0\le\g\le1$,
where $\g=0$ for dust and $\g=1$ for radiation. Substituting in
(\ref{cons}) and integrating, one obtains
\be\lb{ro}
\rho R^{1+\gamma}=M/2\pi,
\ee
with $M$ an integration constant. Substituting again in
(\ref{feq2}), one has
\be\lb{feq4}
\dot R\dot\eta-\Lh R\eta=4GMR^{-\g},
\ee
from which one can easily determine $\y$.
We distinguish three cases: 1) $\L=0$, 2) $\L<0$, 3) $\L>0$.

\subsection{$\L=0$}
There are two possible solutions: either both $R$ and $\y$ are
constant, or $R=At$, with $A=\sqrt a$ and\footnote{ Here and in the
following we choose the origin of time so that it simplifies the
expression of the solutions.}
\bea
&\y={4GM\over A^{1+\g}}\left({t^{1-\g}\over1-\g}-b\right)&\qquad{\rm if}
\ \g\ne1,\nn\\
&\y={4GM\over A^2}(\log t-b)&\qquad{\rm if}\ \g=1,\nn
\eea
with $b$ an integration constant.
In both cases the spacetime is flat. However, if $b>0$, the time-dependent
solutions have a zero of the dilaton at $t_0=[(1-\g)b]^{1/(1-\g)}$,
which we interpret as an initial singularity. These solutions can
therefore be viewed as expanding universes that begin at time
$t_0$ with finite size $At_0$.

\subsection{$\L<0$}
Integrating (\ref{feq1a}), one obtains
\bd
R=A\sin\l t,
\ed
where $\l=\sqrt{-\L/2}$, $A=\sqrt a/\l$. From (\ref{feq4}), one
can write down $\y$ in terms of hypergeometric functions,
\bd
\y=\y_0\cos\l t+{4GM\over A^{1+\g}\l^2}\ \rF\left(\mbox{\small $-\ha\,
,{1+ \g\over2}\,,\ha\,,\cos^2\l t$}\right),
\ed
with $\y_0$ an integration constant. In particular,
\bea
&\y=\y_0\cos\l t+{4GM\over A\l^2}\sin\l t&\qquad{\rm if}\ \g=0,\nn\\
&\y=\y_0\cos\l t+{4GM\over A^2\l^2}\left(1+\cos\l t\,\log\tan{\l t\over2}\right)
&\qquad{\rm if}\ \g=1.\nn
\eea

The metric is that of anti-de Sitter spacetime and describes periodic
solutions. However, for any $\g$, there is a range of values of $\y_0$
for which the dilaton has a zero (corresponding to a physical singularity) at
a finite time $t=t_0$. Such solutions can then be interpreted as universes
which begin expanding with finite initial size at $t_0$ and then recollapse.
Also solutions with both an initial and a final singularity of
dilatonic type may occur for some values of the parameters.

\subsection{$\L>0$}
In this case, the integration of (\ref{feq1a}) yields different results
depending on the sign of $a$. Defining $\l=\sqrt{\L/2}$,
the metric function can assume three qualitatively different forms:
\bea
a<0\qquad&&R=A\ch\l t,\nn\\
a=0\qquad&&R=A\e^{\l t},\nn\\
a>0\qquad&&R=A\sh\l t,\nn
\eea
to which correspond dilaton solutions that are given, respectively, by
\bea
&&\y=\y_0\sinh\l t+{4GM\over\l^2A^{1+\g}}\ \rF\left(\mbox{\small
$-\ha\,,{1+\g\over2}\,,\ha\,,-\sinh^2\l t$}\right),\nn\\
&&\y=\y_0\e^{\l t}-{4GM\over(2+\g)\l^2A^{1+\g}}\e^{-(1+\g)\l t},\nn\\
&&\y=\y_0\cosh\l t-{4GM\over(2+\g)\l^2(A\cosh\l t)^{1+\g}}\
\rF\left(\mbox{\small$1+{\g\over2}\,,{1+\g\over2}\,,2+{\g\over2}\,,
{1\over\cosh^2\l t}$}\right).\nn
\eea
In particular, for $\g=0$, the solutions reduce to
\bea
&&\eta=\eta_0\sh\l t-{4GM\over\l^2A}\ch\l t,\nn\\
&&\eta=\eta_0\e^{\l t}-{2GM\over\l^2A}\,\e^{-\l t},\nn\\
&&\eta=\eta_0\ch\l t+{4GM\over\l^2A}\sh\l t.\nn
\eea
For $\g=1$ one has instead
\bea
&&\eta=\eta_0\sh\l t-{4GM\over\l^2A^2}(1+\sh\l t\,\arctan\sh\l t),\nn\\
&&\eta=\eta_0\e^{\l t}-{4GM\over3\l^2A^2}\e^{-2\l t},\nn\\
&&\eta=\eta_0\ch\l t+{4GM\over\l^2A^2}(1+\ch\l t\,\log\th\l
t/2).\nn
\eea

In all cases, the solutions are locally de Sitter, but have different
global properties. In particular, $R$ has a zero at finite time
if $a>0$, but not in the other cases. Also for $\L>0$
there is a large range of values of $\y_0$ for which all
solutions have a zero of the dilaton at time $t_0$, where $R\ne0$;
the physical behaviour of all of them
is similar, and corresponds to universes
starting at time $t_0$ with a finite size and expanding forever.

\section{Conformal coupling}
We consider now the case in which the matter is linearly coupled
to the dilaton, i.e. $L_M=-\y\r$. This model is invariant under
rescaling of the dilaton, which is fixed up to a multiplicative
constant $\y_0$. The field equations read
\bea
&&\er=\L+16\p G\r,\\
&&-(\na_\mu\na_\nu-g_\mn\na^2)\eta+\Lh g_\mn\eta=8\pi G\y T_\mn.
\eea
Contrary to the model studied in the previous section, one has now
a direct coupling
between the matter density and the curvature of spacetime.

Substituting the ansatz (\ref{ans}) into the field equation, one obtains
\bea
&&{\ddot R\over R}=\Lh+8\p G\r,\lb{eeq1}\\
&&{\dot R\over R}\dot\eta=\Lh\eta+8\pi G\y\r,\lb{eeq2}\\
&&\ddot\eta=\Lh\eta-8\pi G\y p.\lb{eeq3}
\eea
Combining (\ref{eeq1}-\ref{eeq3}), one can check that the
conservation law (\ref{cons}) is still valid and hence also
(\ref{ro}) holds. Eq. (\ref{eeq2}) can then be written as
\be\lb{eeq2a}
\dot R\dot\y=\left(\Lh R+4GMR^{-\g}\right)\y.
\ee
Moreover, using (\ref{ro}), eq. (\ref{eeq1}) can be integrated
once to read
\bea\lb{ceq}
&\dot R^2-\Lh R^2-{8GM\over1-\g}R^{1-\g}=a,&\qquad\se\ \g\ne1,\cr
&\dot R^2-\Lh R^2-8GM\log R=a,&\qquad\se\ \g=1,
\eea
with $a$ an integration constant.
The equations above can be integrated in terms of elementary
functions only when $\g=0$ or when $a=0$ and $\g\ne1$, so in the following
we shall
limit our considerations to these cases. Again, we must distinguish
three possibilities according to the sign of the cosmological constant.

\subsection{$\L=0$}
In this case the integration of (\ref{ceq}) for $\g=0$ gives
\bd
R=2GMt^2-b,
\ed
with $b=a/8GM$.
Integration of (\ref{eeq2a}) gives for the dilaton
\bd
\y=\y_0 t.
\ed
The dilaton is singular at $t=0$. Moreover, if $b<0$, the metric
is always regular, while, if $b\ge0$, a curvature singularity is located at
$t_0=\sqrt{b/2GM}$.
Depending on the value of $b$, the universe begins at $t=0$ with finite size,
or at $t=t_0$ with zero size, and expands forever.

A special solution can be obtained also for $0<\g<1$ if the
integration constant $a$ vanishes. In that case,
\bd
R=\left({2GM(1+\g)^2\over(1-\g)}\right)^{1/(\g+1)}|t|^{2/(\g+1)},\qquad
\y=\y_0|t|^{(1+\g)/(1-\g)}.
\ed
Both the metric function $R$ and the dilaton have a zero at $t=0$, corresponding
to a physical singularity, and grow monotonically with time.

\subsection{$\L<0$}
If $\g=0$, the solution is
\bea
&&R=A\sin\l t\ +4GM/\l^2,\nn\\
&&\y=\y_0\cos\l t,\nn
\eea
where $A=\l^\mo\sqrt{a+(4GM/\l)^2}$. A dilaton singularity
occurs at $t=0$ and a curvature singularity at $t_0=\arccos
(4GM/A\l^2)$, if $A<4GM/\l^2$.
The universe begins expanding at $t=0$ or $t=t_0$ and then recollapses.

If $\g\ne0$ one can find also static solutions.
These have positive $\r$, but negative pressure, namely $\r=-p=-\L/16G\g$.

Other exact solutions can be found  if $a=0$ and $\g<1$. They read
\bea
&&R=\left|{8GM\over(1-\g)\l}\sin{(1+\g)\l t\over2}\right|^{2/(1+\g)},\nn\cr
&&\y=\y_0\left|\cos{(1+\g)\l t\over2}\right|^{2/(1+\g)}\left|\sin{(1+\g)\l
t\over2}\right|^{(1-\g)/(1+\g)}.\nn\cr
\eea
At $t=0$ both the metric and the dilaton are singular. At
$t=\pi/(1+\g)\l$, where the universe reaches its maximum
expansion, the dilaton has a zero, and hence a physical
singularity occurs.

\subsection{$\L>0$}
For positive cosmological constant and $\g=0$, one can obtain three
different solutions
depending on the value of $a$ being lower, equal or greater than
$(4GM/\l)^2$.
Defining $A=\l^\mo\sqrt{|a-(4GM/\l)^2|}$, one has, respectively,

\bea
&&R=A\cosh\l t\ -4GM/\l^2,\nn\\
&&\y=\y_0\sinh\l t.\nn
\eea

\bea
&&R=\l^{-2}(\e^{\l t}\ -4GM),\nn\\
&&\y=\y_0\e^{\l t}.\nn
\eea

\bea
&&R=A\sinh\l t\ -4GM/\l^2,\nn\\
&&\y=\y_0\cosh\l t.\nn
\eea

The first case is similar to the previous ones: a curvature
singularity is present if $A<4GM/\l^2$ (i.e. $a<0$), while the
dilaton is always singular at $t=0$.
In the remaining cases, a curvature singularity occurs at the zero of $R$,
while $\y$ is regular everywhere.
In all cases the universe expands forever.

If one requires positive $\r$, no static solutions exist for $\L>0$.
Special solutions can be found for $a=0$, $\g\ne1$. They read
\bea
&&R=\left|{8GM\over(1-\g)\l}\sinh{(1+\g)\l t\over2}\right|^{2/(1+\g)},\nn\cr
&&\y=\y_0\left|\cosh{(1+\g)\l t\over2}\right|^{2/(1+\g)}\left|\sinh{(1+\g)\l
t\over2}\right|^{(1-\g)/(1+\g)}.\nn\cr
\eea
These solutions are qualitatively similar to those occurring for
$\L=0$, $a=0$. A curvature and a dilaton singularity occur at $t=0$, after
which $R$ and $\y$ grow monotonically.

\section{Particle horizons}
An important property of cosmological models is the existence of
particle horizons, defined as the location of the most distant
place from which a light ray can have reached us since the beginning
of the universe. It is easy to see that its distance is proportional
to the integral
\be\lb{hor}
\int_{t_0}^t {dt'\over R(t')}
\ee
where $t_0$ is the initial time.
If this integral is infinite, no particle horizon is present.

For our models, one must distinguish two cases: when the initial singularity
is a dilatonic one, the metric is regular at $t_0$ and therefore
the integral (\ref{hor}) cannot diverge, and a particle horizon
always exists.
When one has an initial curvature singularity, instead, a computation of
the integral (\ref{hor}) shows that in all cases, except the de Sitter
solutions $\L>0$, $a<0$ of section 2.3, it diverges at the initial singularity,
and hence no particle horizon is present.

\section{Final remarks}
We have studied two-dimensional cosmologies in the context of the
Jackiw-Teitelboim model, in the case of minimally coupled and
conformally coupled matter. All solutions present initial singularities.
However, these can be either of metric or dilatonic nature.
In the latter case, the universe can have
a finite size at its beginning. The universe expands forever or
recollapses depending on the value of the cosmological constant.
In general, a particle horizon only exists in the case of dilaton
singularities.

\end{document}